\documentclass{article}
\usepackage{spconf,amsmath,graphicx}
\usepackage{amsmath}
\usepackage{multirow}
\usepackage{graphicx} 
\usepackage{amsthm}
\usepackage{algpseudocode}
\usepackage{threeparttable}
\usepackage{algorithm}
\usepackage{epstopdf}
\usepackage{latexsym,bm,amsmath, amssymb,amsfonts}%
\usepackage{enumitem}
\usepackage{booktabs} 
\usepackage{color}
\usepackage[format=plain]{caption}
\usepackage{enumitem}

\definecolor{orange}{RGB}{255,107,0}

\title{Counterfactual Regret Minimization for Anti-jamming Game of Frequency Agile Radar}
\name{Huayue Li$^\dagger$, Zhaowei Han$^\dagger$, Wenqiang Pu$^\star$, Liangqi Liu$^\dagger$, Kang Li$^\ddag$, Bo Jiu$^\ddag$\thanks{Huayue Li and Zhaowei Han contribute equally to this work. }}
\address{$^\dagger$The Chinese University of Hong Kong, Shenzhen, China\\
$^\star$Shenzhen Research Institute of Big Data, China\\
$^\ddag$National Laboratory  of  Radar  Signal  Processing,  Xidian  University, Xi’an, China}
\begin{document}
\ninept
\maketitle
\begin{abstract}
The competition between radar and jammer is one emerging issue in modern electronic warfare, which in principle can be viewed as a non-cooperative game with two players. In this work, the competition between a frequency agile (FA) radar and a noise-modulated jammer is considered. As modern FA radar adopts coherent processing with several pulses, the competition is hence in a multiple-round way where each pulse can be modeled as one round interaction between the radar and jammer. To capture such multiple-round property as well as imperfect information inside the game, i.e., radar and jammer are unable to know the upcoming signal, we propose an extensive-form game formulation for such competition. Since the number of game information states grows exponentially with respect to number of pulses, finding Nash Equilibrium (NE) strategies may be a computationally intractable task. To effectively solve the game, a learning-based algorithm called deep Counterfactual Regret Minimization (CFR) is utilized. Numerical simulations demonstrates the effectiveness of deep CFR algorithm for approximately finding NE and obtaining the best response strategy.

\end{abstract}

\section{Introduction}

Electronic counter-countermeasures (ECCM) is one emerging issue in radar signal processing area. Many signal processing techniques \cite{zhou2017parameter,4472184,8378682} were proposed in the past and these techniques can be regarded as \textit{passive} approaches which try to eliminate the jamming signals after the radar has been jammed. The anti-jamming performance of separating target signal from jammed signal is usually limited. Instead, \textit{active} approaches which take proper strategy over parameters of transmit signal in advance to avoid being jammed potentially admits better performance. Frequency agile (FA) radar which is capable of changing the carrier frequency of the transmit signal is one representative radar system to realize active anti-jamming approach. The key of success anti-jamming is the strategy for carrier frequency selection. 

Recently, several strategy design methods based on reinforcement learning (RL) are studied in the literature~\cite{sp_underreview,9266402,9114797,9322520}, i.e., the jammer is modeled as the environment and the radar is the agent who makes sequential decisions based on interactions with the jammer. The work in~\cite{sp_underreview} considers carrier frequency selection of the FA radar and a deep RL approach is proposed. In~\cite{9266402}, Q-learning is proposed for selecting both the carrier frequency and pulse width. In~\cite{9114797}, the anti-jamming problem of the FA radar is modeled as a partially observable Markov decision process and a deep Q-network learning approach is proposed. Modeling the jammer as a stationary environment, i.e., the jamming rule is fixed, may not be effective for combating jammer with time-varying rule. Recent work~\cite{li2021robust} proposes a robust anti-jamming strategy design to deal this issue. However, the aforementioned  single-agent  RL based work ignores  the  learning  ability  of  the  jammer, which  may limit their practical usage. Instead, modeling the competition between the radar and jammer as a game can capture the learning ability of both the radar and jammer. As modern FA radar usually adopts coherent processing with several pulses, classical norm-form game can not describe the interaction among multiple pulses and possible imperfection information in the competition. 

In this work, we introduce the extensive-form game (EFG) to model the competition between the radar and jammer. EFG can be expressed by a game tree, which records the information of the interaction for multiple pulses. Since the size of EFG grows exponentially with respect to number of pulses, finding Nash Equilibrium (NE) strategies may be a computationally intractable task. To effectively solve the game, a learning-based algorithm called deep Counterfactual regret minimization (CFR) is utilized. Simulations demonstrates the effectiveness of deep CFR algorithm for finding NE in self-play setting (both the radar and jammer play with deep CFR algorithm) and obtaining best response strategy against other one's stationary strategy.
\section{System Model}
\subsection{Signal Model}\label{subsec:sigmodel}
Consider a FA radar equipped with a transmitter and a receiver antenna array. The transmitted signal at radar is $s(t) = \sum\nolimits_{m=1}^M {s}_m(t - (m-1)T)$, where $T$ is pulse repetition time, $M$ is the number of pulses, and ${s}_m(t)$ is the transmitted signal at pulse $m$. Each ${s}_m(t)$ contains $K$ sublupses and ${s}_m(t)$ is modeled as 
\begin{equation*}
s_m(t)=a(t)\sum\nolimits_{k=1}^K \text{rect}((t - k T_c)/T_c)\text{exp}(j 2 \pi f_{k}^m t),
\end{equation*}
where $a(t)$ is the complex envelope, $T_c<T$ the duration time of subpulse, $\text{rect}(t)$ stands for the rectangle function which is equal to one if $t$ belongs to $[0, 1]$ and zero elsewhere. Parameter $f_k^m\in\mathcal{F}=\{ \bar{f}_1,\ldots,\bar{f}_N \}$ is the carrier frequency and $\mathcal{F}$ denotes the available carrier frequency set. 

\begin{figure}
\centering
\includegraphics[width=1\linewidth]{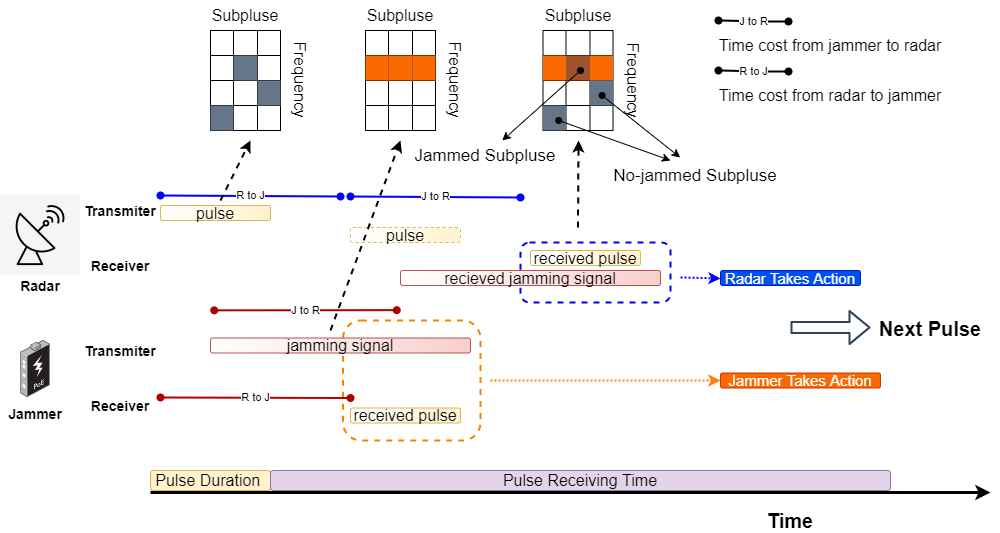} 
\caption{Illustration of the signal model in one pulse.}
\label{fig:onepulse}
\end{figure}

In the environment, there is a jammer transmitting signal $u(t) = \sum\nolimits_{m=1}^M u_m(t-(m-1)T)$, where $u_m(t)$ is the jamming signal for the $m$-th pulse. And  
\begin{equation*}
    u_m(t) = \text{rect}(t/T_J) v_m(t)\text{exp}(j 2 \pi f_m t),
\end{equation*}
where $f_m\in\mathcal{F}$ is the carrier frequency of jamming signal, $v_m(t)$ is the jamming signal envelope, $T_J> 0$ is the duration time of jamming signal. An illustration of the signal model for one pulse is given in Fig. \ref{fig:onepulse}.

In this work, we consider both the radar and jammer are able to change their carrier frequency of each pulse. The general goal of the radar is to avoid its transmitted signal $s(t)$ being jammed by $u(t)$ in frequency domain and the jammer has an opposite goal. Notice that the above signal models are introduced for clearly explaining the competition from signal level. As both the radar and jammer can receive its opponent's signal from which the frequency information can be extracted. The competition thus can be `abstracted' as a game for frequency selection. In the rest part of this paper, we purely consider the radar and jammer play game in frequency domain by using concepts in game theory.

\subsection{Action Sets for Radar and Jammer}\label{subsec:actset}
Based on the above signal models, the action set of the radar and jammer for each pulse are as follow:

\noindent \textbf{Radar's Action Set}: At each pulse, the radar needs to decide carrier frequency for each subpulse, i.e., $f_k^m,\forall k$. Hence, by independence of each subpluse, its action set is a Cartesian product of $\mathcal{F}$ of order $K$, denoted as 
$\mathcal{A}_R=\mathcal{F}\times \mathcal{F}\times \ldots\mathcal{F}=\mathcal{F}^K.$

\noindent \textbf{Jammer's Action Set}: The jammer considered in this work is a self-protection jammer which transmits noise modulated jamming  signal and works in different mode. In particular, the jammer can adopt three modes: spotting jamming mode where the jammer transmits narrow band noise signal at carrier frequency $f_m\in\mathcal{F}$, barraging jamming mode where the jammer transmits wide band jamming signal centered at a pre-determined carrier frequency $f_0$, or reacting jamming mode where the jammer intercept the first subpluse of the received radar pulse and then transmits noise signal at the intercepted carrier frequency, i.e., $f_m=f_1^m$. Hence, denote symbols `$\textrm{Ba}$' and `$\textrm{Ra}$' as the barraging mode and reacting mode respectively, the jammer's action set can be represented as 
$\mathcal{A}_J=\mathcal{F}\cup \{ \textrm{Ba},\textrm{Ra}\}.$


\section{Extensive Form Game}
\label{sec:gamemodel}

In this section, we introduce a game form called \textit{extensive form game} to model the competition between the radar and jammer. This game is multiple-round, where each round corresponds to one pulse and there are $M$ rounds (pulses) in total. At the beginning of each round, both the radar and jammer simultaneously\footnote{From game perspective, the radar and jammer simultaneously take action though they may not physically take action at the same time.} decide and take their action. This way, the opponent's action is unknown until the end of current round. This brings \textit{imperfect information} for decision making. In game theory, extensive form game is a tree-based formalism used to describe a large class of imperfect information games. Formally speaking, an extensive form game $\mathcal{G}$ is a tuple $\langle H,Z ,P,p,u,I,\sigma_c \rangle$ and detailed explanation of each notation are as follow:

\begin{itemize}[leftmargin=10pt]
    \item \textbf{Player Set $\mathcal{P}$:} The set of all players acting in the game is denoted as $\mathcal{P}$. In the considered anti-jamming game, $\mathcal{P}$ consists of $2$ players, i.e., a radar player and a jammer player. 
    \item \textbf{Action Set $\mathcal{A}$:} All possible actions that players can take is collected as a finite set $\mathcal{A}$. For each player, its legal action set is a subset of $\mathcal{A}$ depends on the player and current game state. In the considered anti-jamming game, we have $\mathcal{A}=\mathcal{A}_R\cup \mathcal{A}_J$ where $\mathcal{A}_R$ and $\mathcal{A}_J$ are action sets defined in Section~\ref{subsec:actset}. 
    \item \textbf{History Set $\mathcal{H}$:} Each history $h\in\mathcal{H}$ is defined as a sequence of actions that were taken by players from start of the game. Denote $a_R^m\in\mathcal{A}_R$ and $a_J^m\in\mathcal{A}_J$ as the action taken by the radar and jammer for pulse $m$ respectively, then history for the radar and jammer at pulse $m$ is denoted as $h_R^m$ and $h_J^m$ respectively,
    \begin{align*}
        h_R^m&=(a_R^1,a_J^1,a_R^2,\ldots,a_R^{m-1},a_J^{m-1},a_R^{m},\emptyset,\ldots,\emptyset)\in\mathcal{H},\\
        h_J^m&=(a_R^1,a_J^1,a_R^2,\ldots,a_R^{m},a_J^{m},\emptyset,\ldots,\emptyset)\in\mathcal{H},
    \end{align*}
    where both $h_R^m$ and $h_J^m$ are of length $2M$ and $\emptyset$ represents empty action just for notation convenience. 
    \item \textbf{Terminal Histories $\mathcal{Z}$:} Each terminal history $z\in \mathcal{Z} \subseteq \mathcal{H}$ represents a complete game play. For the considered anti-jamming game, we have $z=h_R^M$.
    \item \textbf{Utility Function $u(z)$:} At end of the game, $|\mathcal{P}|$ players receive their corresponding utility $u(z):\mathcal{Z}\mapsto \mathbb{R}^{|\mathcal{P}|}$. For the considered anti-jamming game, the probability of detection (PD)[] is used as utility which can be calculated from terminal history $h_J^M$. 
    \item \textbf{Player Identity Function $p$:} Each non-terminal history corresponds to one player taking action and function $p:\mathcal{H}/\mathcal{Z} \mapsto P$ identifies this acting player.
    \item \textbf{Information State Set $\mathcal{S}$:} The concept of information set is used to capture imperfect information in the game. Each information state $s\in\mathcal{S}$ consists of at least one history $h\in s$ and for all $h,h^\prime \in s$, we have $p(h)=p(h^\prime)$. During the game play, player only knows which information state $s$ they are staying at but can not distinguish history $h\in s$. Therefore, $\mathcal{S}$ is a partition of $\mathcal{H}$ and different game rule leads different partition form. For the anti-jamming game at the $m$-th round, we treat the radar as the `first' player such that all possible $h_R^m$ form one information state since the jammer can not observe the radar's action unless it plays as the `second' player. As both the radar and jammer can observe their opponents' actions at the end of $m$-th round, each $h_J^m$ forms one information state. 
    \item \textbf{Behavior Strategy $\pi_p(s)$:} At each information state $s$, let $p=p(h),h\in s$, player $p$ takes action according to behavior strategy $\pi_p(s)\in \Delta(\mathcal{A}_s)$, where $\mathcal{A}_s\subseteq \mathcal{A}$ is the legal action set at $s$ and $\Delta(\cdot)$ denotes probability simplex over a discrete set. For the anti-jamming game, we simplify notation as $\pi_{R}(s)\in\Delta(\mathcal{A}_R)$ and $\pi_J(s)\in\Delta(\mathcal{A}_J)$ for the radar and jammer respectively. 
\end{itemize}
Based on the above concepts, an illustration the anti-jamming game tree is given in Fig. 2. 
\begin{figure*}
\centering
\includegraphics[width=1\linewidth]{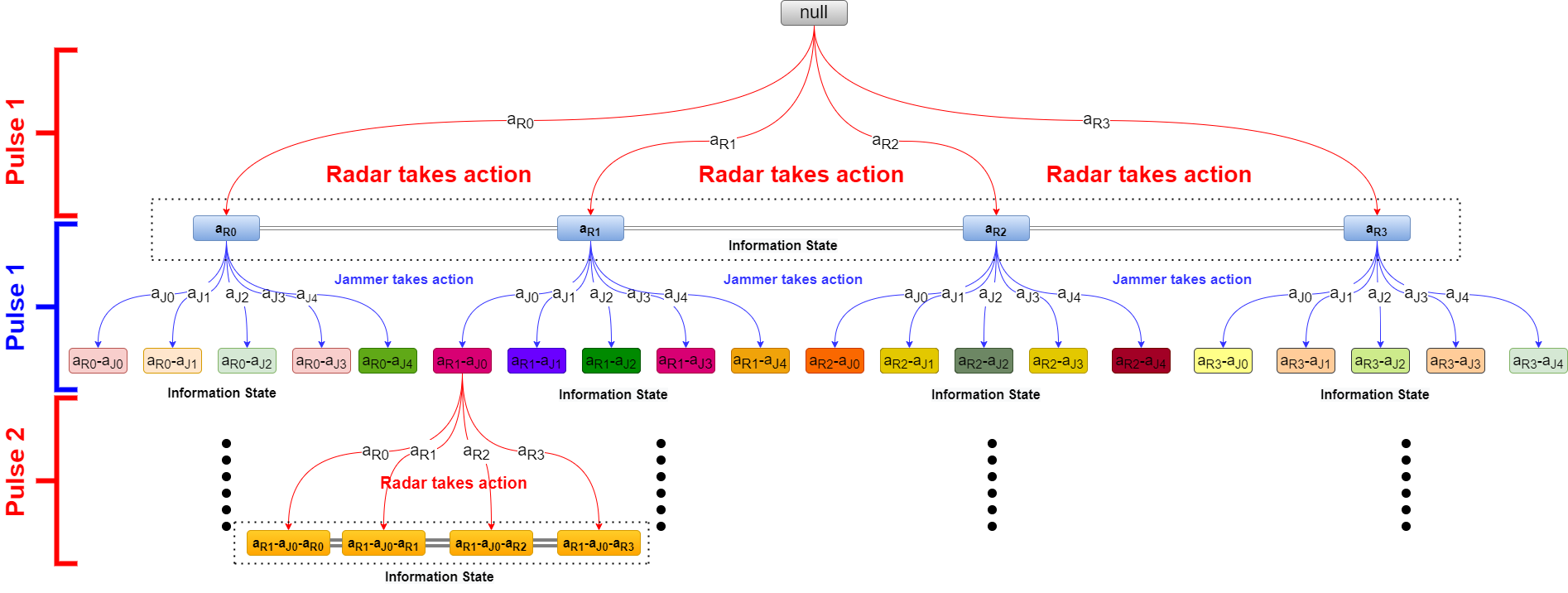} 
\caption{Game tree representation of the extensive-form game between the radar and jammer. Blocks of different colors represent different information states and blocks that are covered by line of dashes belongs to the same information state. $a_{R}$ (total $4$ actions) and $a_{J}$ (total $5$ actions) denote actions of the radar and jammer respectively.}

\label{fig:info_set}
\end{figure*}

\section{Deep Counterfactual Regret Minimization}
In game literature, one effective algorithm for solving extensive form game is counterfactual regret minimization (CFR) introduced by Zinkevich \emph{et al}. Here after we briefly revisit this algorithm. 

Recall the behavior strategy $\pi_p(s)$ introduced in Section~\ref{sec:gamemodel}, we use $\bm{\pi}=\{\pi_p(s)\mid \forall s\in\mathcal{S},p\in\mathcal{P} \}$ to denote the strategy profile. The key concept in CFR is the counterfactual value at information state $s$ (where player $p$ takes action), which is defined as 
$$
v^{\bm{\pi}}(s)=\sum_{h\in s} \sigma_{-p}^{\bm{\pi}}(h)\sum_{z\in\mathcal{Z}}\sigma^{\bm{\pi}}(h,z)u_p(z),
$$
where $\sigma_{-p}^{\bm{\pi}}(h)$ is the probability of reaching $h$ contribute by $\bm{\pi}$ and $\sigma^{\bm{\pi}}(h,z)$ is the probability of reaching $z$ given $h$.
Precise definition of $\sigma_{-i}^{\bm{\pi}}(h)$ and $\sigma^{\bm{\pi}}(h,z)$ requires complicated notations and we refer~\cite{brown2019deep} for more details. This way, we can define counterfactual value at information state $s$ with respect to an action $a$ as 
$$
v^{\bm{\pi}}(s,a)=\sum_{h\in s} \sigma_{-p}^{\bm{\pi}}(h)\sum_{z\in\mathcal{Z}}\sigma^{\bm{\pi}}(ha,z)u_p(z).
$$
Based on the counterfactual values $v^{\bm{\pi}}(s,a),\forall s,a$, CFR updates $\bm{\pi}$ in an iterative way. At the $t$-th iteration, the instantaneous regret with $\bm{\pi}^t$ for action $a$ at information set $s$ is 
$$
r^t(s,a)=v^{\bm{\pi}^t}(s,a) - v^{\bm{\pi}^t}(s).
$$
Correspondingly, the counterfactual regret at iteration $t$ is
$$
R^t(s,a)=\sum\nolimits_{t^\prime=1}^t r^{t^\prime}(s,a).
$$
Comparing with standard definition of regret, counterfactual regret $R^t(s,a)$ is not the `real' regret since it ignored the probability of reaching $s$ that contribute by player $p$ with strategy profile $\bm{\pi}$. CFR updates $\bm{\pi}^{t+1}$ as 
$$
\bm{\pi}^{t+1}(s,a)=\frac{[R^t(s,a)]_+}{\sum_{a\in\mathcal{A}(s)}[R^t(s,a)]_+},
$$
where $[r]_+=\max\{0,r \}$.

By iteratively traversing the game tree, CFR is able to converges to an equilibrium \cite{burch2018time}. Note that for the anti-jamming game of interests, the number of information states grows exponentially with the number of pulses, ordinary CFR is not computationally tractable. Recent work \cite{brown2019deep} incorporates deep neural network in CFR to accelerate the game solving process, named Deep CFR. With an advantage network and a strategy network, Deep CFR is 
demonstrated being able to find NE in poker games, see details in \cite{brown2019deep}. This work will use Deep CRF to solve the anti-jamming game.

\section{Simulation Results}
\vspace{-0.5em}
Based on EFG model introduced in Section~\ref{sec:gamemodel}, some preliminary simulations are conducted to evaluate the performance of Deep CFR.
\vspace{-0.5em}
\subsection{Setups}
\begin{figure}[ht]
\flushleft
\includegraphics[width=0.48\textwidth,height=0.38\textwidth]{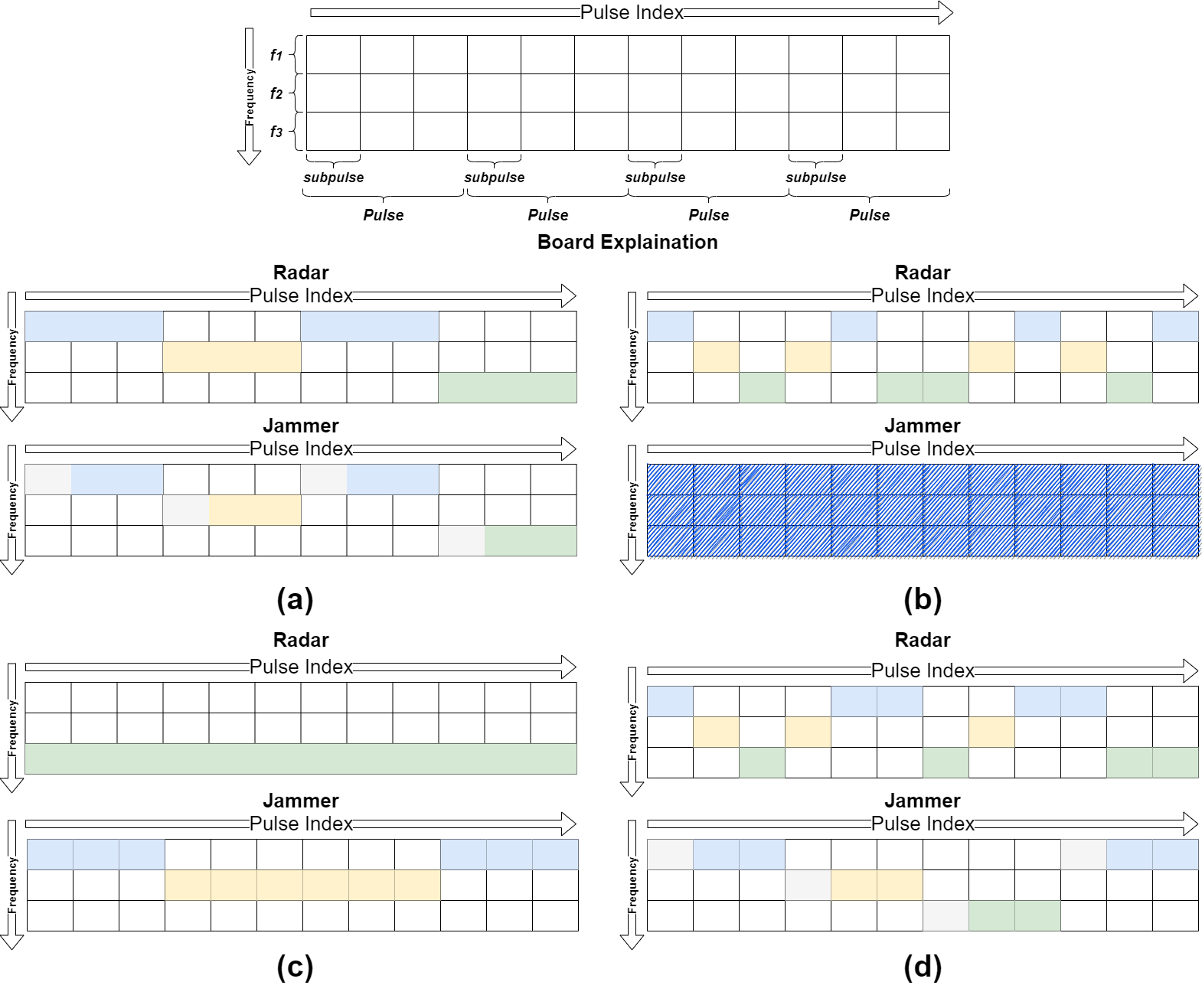}
\caption{Game board explaination and examples of information states. Colored blocks stand for selected frequencies. The block in grey and corrugated blue represents the `Ra' and `Ba' actions respectively}
\label{fig:fixed_policy_cases}
\end{figure}

\vspace{-0.5em}
The anti-jamming game with $M=4$ pulses is considered. Each pulse contains $K=3$ subpulses and the frequency set $\mathcal{F}$ contains $3$ frequencies, i.e., $\mathcal{F}=\{f_1,f_2,f_3 \}$. Action sets $\mathcal{A}_R$ and $\mathcal{A}_J$ are specified accordingly. The game is zero-sum and the probability of detection (PD) is used as the utility function. System parameters of the radar and jammer are summarized in Table~\ref{tab:pd}. For game simulation, the specified anti-jamming game is built based on an open library called OpenSpiel~\cite{lanctot2019openspiel} which is a collection of environments and algorithms for research in reinforcement learning and planning in games. In particular, we embed the information states of the anti-jamming game by using the one-hot method. A tensor board of size $|\mathcal{F}|\times MK\times 2=3\times 12\times 2$ is used to represent the information state (see illustrations in Fig.~\ref{fig:fixed_policy_cases}), where the total number of terminal states is $|\mathcal{Z}|=27^4\times 5^4$. In OpenSpiel, the game play is modeled in a sequential manner with unknown tokens. Radar and jammer take actions sequentially, i.e., radar is the first player, but the jammer won't be able to "see" the radar's action before the end of current round. This captures the imperfection information of the game. Two algorithms, Deep Q-learning (DQN)~\cite{moreno2019performing} and neural fictitious self play (NFSP) \cite{heinrich2016deep} are used as benchmarks. We use 8 layers Multilayer Perceptron (MLP)\footnote{MLP Size: $288\times1024\times512\times512\times256\times256\times128\times128\times32$.} as our policy network and 7 layers MLP\footnote{MLP Size:  $288\times 1024\times512\times256\times256\times128\times128\times32$.} as our advantage network (described in Section 4) in Deep CFR. In DQN and NFSP, we use the same 8 layers MLP. Each layer of MLP execpt last output layer in above algorithms uses rectified linear unit (ReLU) as the activation function. The last output layer for MLP uses Softmax as the activation function.  
\begin{table}
\centering
\caption{System Parameters}
\label{tab:pd}
\scalebox{0.9}{
\begin{tabular}{|c|c|}
\hline
		\cline{1-2}
		\textbf{System Parameter} & \textbf{Value}\\
		\cline{1-2}
        RCS of $f_1,f_2,f_3$ & 15,3,1\\
        \cline{1-2}
        False Alarm Rate& $10^{-4}$\\
        \cline{1-2}
        Bandwidth of Subpulse & $2$MHz\\
        \cline{1-2}
        Noise Bandwidth& $500$MHz\\
        \cline{1-2}
        Power of Subpulse& $30$kW\\
        \cline{1-2}
        Radar Antenna Gain & $30$dB\\
        \cline{1-2}
        Range between Radar and Jammer& $100$km\\
        \cline{1-2}
        Power of Jammer& $10$W\\
        \cline{1-2}
        Jammer Antenna Gain & $6$dB\\
\hline
\end{tabular}}
\end{table}

\subsection{Performance on Achieving Nash Equilibrium}
\vspace{-0.5em}

\begin{figure}[ht]
\centering
\includegraphics[width=0.8\linewidth]{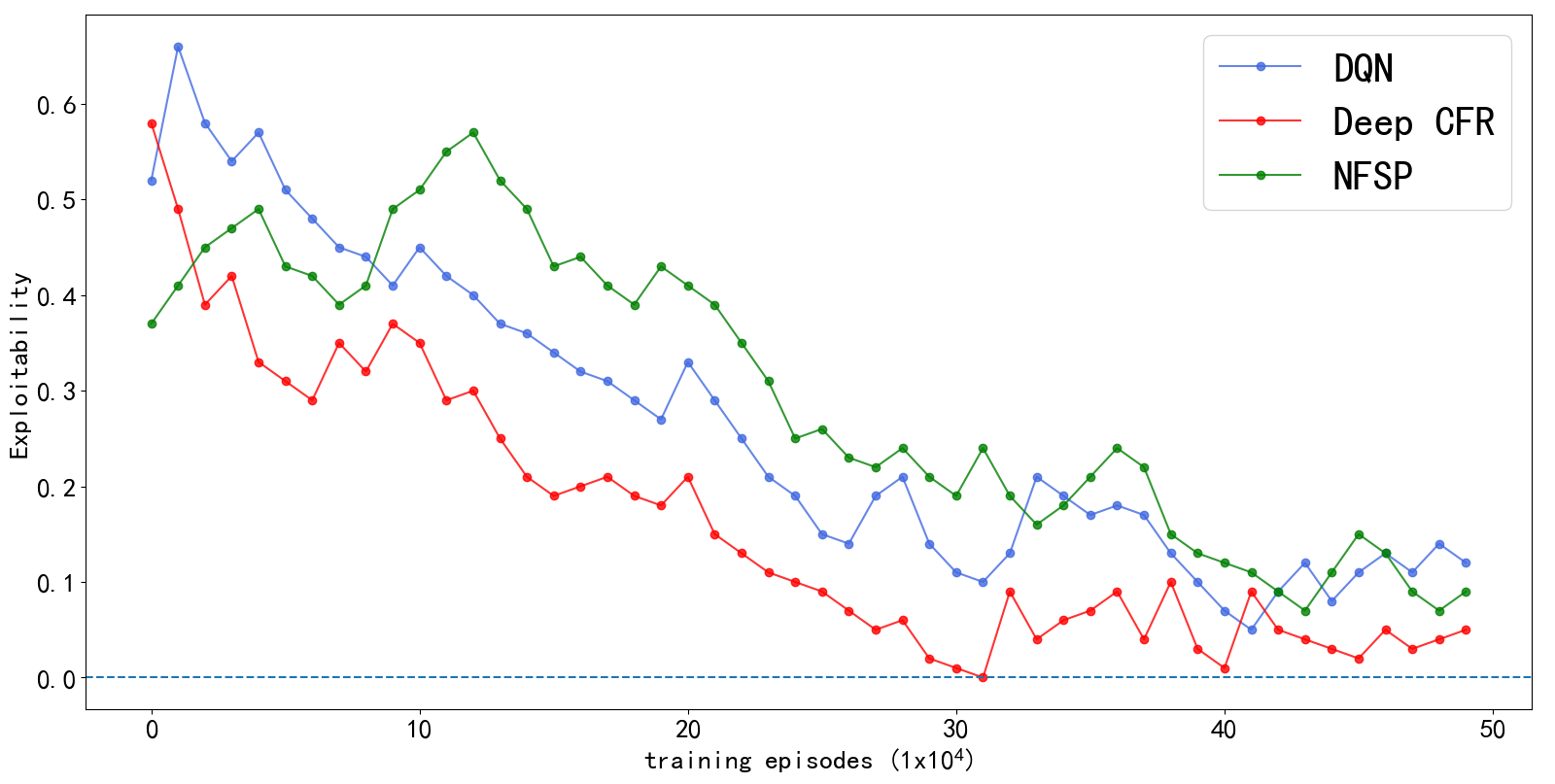} 
\caption{Convergence curves of exploitability.}
\label{fig:Exploitability}
\end{figure}

We first evaluate the performance for achieving Nash equilibrium (NE).  Exploitability \cite{zinkevich2007regret} which describes how far a policy profile $\bm{\pi}$ from NE is used as the performance metric. Both the radar and jammer play the same algorithm and the exploitability curves are compared in Fig.~\ref{fig:Exploitability}. Results show that Deep CFR has faster convergence than the other two baselines. We also note that DQN may not always converge in our simulation, since DQN in general can not be guaranteed to converge in self-play setting.

\begin{figure*}
\centering
\includegraphics[width=1\linewidth]{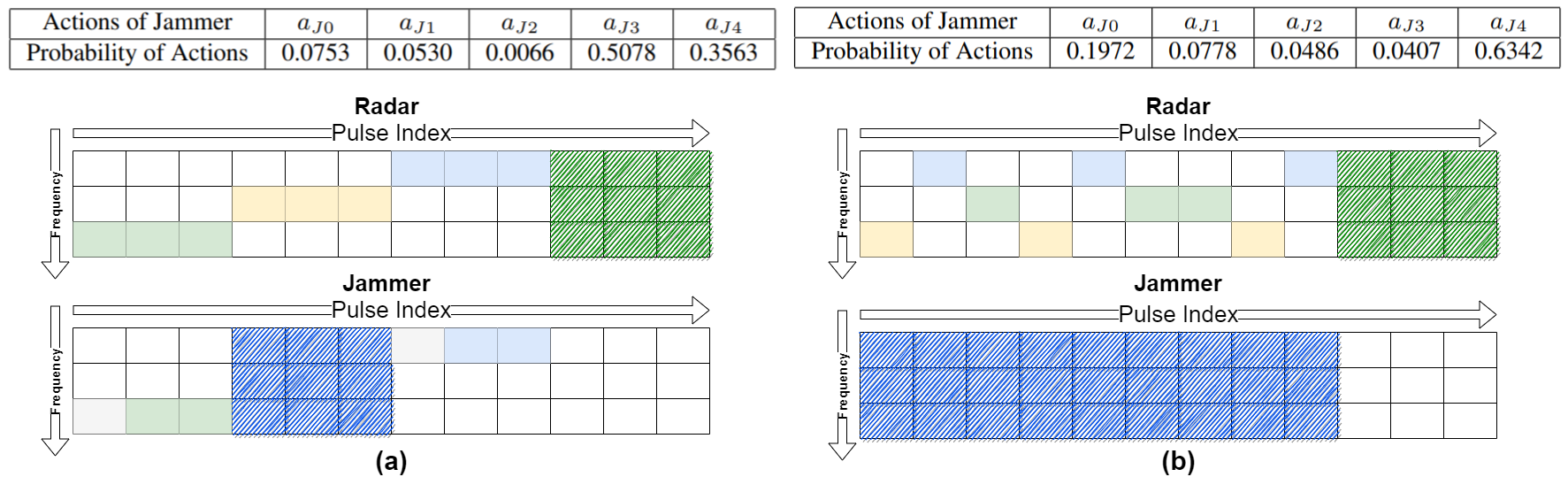} 
\caption{Two states in cases (a) and (b). For radar, the block in corrugated green refers to the unknown token. Two tables show the policy of the jammer in cases (a) and (b) correspondingly. In the  tables, $a_{J0}$-$a_{J2}$ corresponds to frequencies in $\mathcal{F}$, $a_{j_3}$ refers to `Ra' action and $a_{j_4}$ refers to `Ba' action.}
\label{fig:final_result_analysis}
\end{figure*}

\subsection{Performance on Learning Strategies}
\vspace{-0.5em}
Achieving NE is not the only purpose in the anti-jamming problem, we also intend to understand the learning ability of algorithms, i.e., agents are capable of reaching high rewards in different settings. Therefore, we design four special cases (as illustrated in Fig.~\ref{fig:fixed_policy_cases}) to test the learning ability of different algorithms. In case (a) and (b), we treat the radar as the environment and the jammer as the agent who tries to find the best policy to fight against the radar. More specifically, in case (a), radar follows the principle of selecting the same frequency for each subpulse in one pulse. On the contrary, in case (b), the radar follows the principle of never selecting the same frequency in one pulse. For cases (c) and (d), we treat the jammer as the environment instead and the radar learns to fight against the jammer. In case (c), jamming only occurs in two frequencies, i.e., $\mathcal{A}_J=\{f_1,f_2 \}$. In case (d), the jammer is only allowed to take $\textrm{Ra}$ action instead of any other jamming actions. Overall, we expect the agent can learn strategy that achieve higher reward.

In Table~\ref{tab:ul}, the achieved PD after $5\times 10^5$ training episodes are compared. For cases (a) and (b), the lower PD indicates the better performance that the jammer achieved. For cases (c) and (d), the higher PD implies the better performance that the radar achieved. In short, comparing the two algorithms, i.e., Deep CFR and NFSP, which are designed for solving EFG, Deep CFR has similar or better performance. DQN as a single-agent algorithm achieves better performance in cases (b) and (c). Finally, two representative states in cases (a) and (b) are illustrated in Fig.~\ref{fig:final_result_analysis}, where the learned policies of Deep CFR at these states are listed in the corresponding tables. In case (a), the jammer is preferred to choose `Ra' action since the radar always choose the same frequency in one pulse. In case (b), the jammer is preferred to choose `Ba' action since the radar uses different frequency in one pulse.

\vspace{-0.5em}
\begin{table}[H]

\centering
\caption{Achieved utilities in difference cases.}

\scalebox{0.9}{
\begin{tabular}{|c|c|c|c|c|}
\hline
		\multirow{2}{*}{case a}& Algorithms & Deep CFR & DQN & NFSP\\
		\cline{2-5}
		~& Utility & \textbf{0.7832} &0.8669   & 0.8708\\
		\cline{1-5}
        case b&Utility & 0.8851 & \textbf{0.6793} & 0.9073\\
		\cline{1-5}
		case c&Utility & 0.9382 &\textbf{0.9557}   & 0.9335\\
\cline{1-5}
case d&Utility & \textbf{0.9612} & 0.8866   &  0.9605\\
\hline
\end{tabular}
\label{tab:ul}
}
\end{table}

\section{Conclusion}
\vspace{-0.5em}
In this paper, extensive form game is introduced to model the competition between a FA radar and a self-protection jammer. Practical issues like multiple-round interactions and imperfect information are captured by the game model. To solve the game, Deep CFR is utilized to approximately solve the game and some preliminary simulations are conducted to demonstrate its effectiveness. EFG for modeling other related anti-jamming scenarios would be interesting to explore in the future. Learning algorithms for solving the game, e.g., regret matching type (Deep CFR) and self-play type (NFSP), are also important research directions, especially for problem with large size which is the case of many practical anit-jamming scenarios.


\newpage
\bibliographystyle{ieeetr}
\bibliography{test}

\end{document}